\documentclass[]{aa}  

\usepackage{graphicx}
\usepackage{ulem}
\usepackage{hyperref}
\usepackage{txfonts}
\usepackage{todonotes}

\hypersetup{colorlinks=true, linkcolor=blue, citecolor = blue, urlcolor=cyan }
\defcitealias{soyuer2021}{S21}
\begin{document} 

  \title{Prospects for a local detection of dark matter \\with future missions to Uranus and Neptune}

  \author{Lorenz Zwick\thanks{zwicklo@ics.uzh.ch}
          \and
          Deniz Soyuer
          \and
          Jozef Bucko
          }

  \institute{Center for Theoretical Astrophysics and Cosmology, Institute for Computational Science, University of Zurich,\\
             }

  \date{Received: 11th, 04, 2022. Accepted: 14th, 04, 2022}


 
  \abstract
   {}
   {We investigate the possibility of detecting the gravitational influence of dark matter (DM) on the trajectory of prospective Doppler-ranging missions to Uranus and Neptune. In addition, we estimate the constraints such a mission can provide on modified and massive gravity theories via extra-precession measurements using orbiters around the ice giants.}
   {We employd Monte Carlo-Markov Chain methods to reconstruct fictitious spacecraft trajectories in a simplified solar system model with varying amounts of DM. We characterise the noise on the Doppler link by the Allan deviation $\sigma_{\rm A}$, scaled on the Cassini-era value of $\sigma^{\rm{Cass}}_{\rm A}= 3 \times 10^{-15}$. Additionally, we compare the precision of prospective extra-precession measurements of Uranus and Neptune with the expected rates from simulations in the context of modifications to the inverse square law.
   }
   {We estimate that the prospective mission will be sensitive to DM densities of the order of $\rho_{\rm{DM}} \sim 9 \times 10^{-20} \, (\sigma_{\rm A}/\sigma_{\rm A}^{\rm{Cass}}) $ kg/m$^3$, while the $1\sigma$ bound on the expected galactic density of $\rho_{\rm{DM}} \sim 5 \times 10^{-22}$ kg/m$^3$ decreases as $1.0 \times 10^{-20} \, (\sigma_{\rm A}/\sigma^{\rm{Cass}}_{\rm{A}})^{0.8}$ kg/m$^3$.  An improvement of two to three orders of magnitude from the baseline Allan deviation would guarantee a local detection of DM. Only a moderate reduction in ranging noise is required to rule out Milgrom's interpolating function with solar system based observations, and improve constraints the graviton mass beyond current local-based or gravitational wave-based measurements. Our analysis also highlights the potential of future ranging missions to improve measurements of the standard gravitational parameters in the solar system.
   }
   {We believe that a ranging mission to Uranus and Neptune also presents an unique opportunity for non-planetary science. The noise improvements required to guarantee a local detection of dark matter in the early 2040s are realistic, provided they become one of the priorities during mission development.}


   \maketitle
%

\section{Introduction}
\label{sec:intropy}
Uranus and Neptune are notoriously underexplored compared to the other solar system giants, and the interest for new in situ missions has been steadily increasing over the past few years \citep[e.g.][]{hofstadter, fletcher, fletcher2,helled, quest, simon2020, kollmann, odyssey}. In addition to the primary science objectives listed in these publications, prospective missions to the ice giants also present an exceptional opportunity for non-planetary science. In this work, we explore the constraints such a mission can provide on the local dark matter (DM) density and on various modified gravity (MG) theories.
As a spacecraft cruises interplanetary space, its motion is affected by the gravitational field of any massive object in the solar system. Moreover, subtle effects such as radiation pressure and the galactic tidal field can produce small accelerations which modify the trajectory over long periods. In the same vein, the presence of DM would produce a radial force proportional to the enclosed mass within the spacecraft's orbit.
If the velocity and trajectory of the spacecraft are tracked with sufficient precision, it is possible to reconstruct a model of all forces acting on the spacecraft and to attribute the additional radial component to the presence of DM, or the effect of a MG scenario. 

Ranging satellites and the Earth are in constant communication through a radio link, which is commonly around the frequency of 32 GHz. Any Doppler shift present in the link is a measure of the spacecraft's radial velocity. Earth's rotation around the Sun introduces a parallax motion which can be used to trace the tangential velocity, allowing for a precise reconstruction of spacecraft trajectories. However, the precision with which the trajectory can be reconstructed is limited by the noise on the Doppler link, which are dominated by astrophysical (such as plasma scintillation), tropospheric and mechanical sources \citep{armstrong}. For our purposes, it is convenient to parametrise this noise via the Allan deviation, a measure of the frequency-averaged stability of the Doppler link. A benchmark measure for modern space missions is to achieve an Allan deviation of $\sigma_{\rm{A}} \leq 3\times 10^{-15}$, comparable with the Cassini probe launched in 1997. This allows for measurements of the radial velocity with a precision of the order of 0.1 mm/s, which is a requirement for precise interplanetary trajectories \citep[see e.g.][]{Deboy2005}.
While further improvements in the Allan deviation are not strictly required for interplanetary navigation, they would be crucial for detecting the minute changes in velocity caused by DM or MG.

The tracking of a ranging spacecraft's trajectory is not the only way the local dark sector can be probed. Several publications have investigated how observational bounds on the precession rate of solar system planets constrain the local DM density and the length scales of MG \citep{sereno_dm,Willgravitoni,local_dark}. The tightest constraints arise from the motion of Mars, since its orbital precession is known with extremely high precision from Doppler ranging measurements of the several artificial satellites that orbit it \citep[][]{pitjeva_2013,Willgravitoni}. However, these constraints are still several order of magnitudes above what is required to achieve predicted DM densities of $\rho_{\rm DM}\sim 5\times 10^{-22}$ kg/m$^3$ from galactic profiles \citep{Weber2010, kafle14}.
The presence of DM should influence the ice giants more strongly because of the larger enclosed volume within their orbits. However, the precession rates of the outermost planets are currently only constrained by astrometric observations, a method that is much less precise than Doppler ranging. As has been noted in \cite{sereno_dm}, a ranging mission to Uranus or Neptune has the potential to improve current measurements on the ice giant precession rates by several orders of magnitude, which would provide stronger constraints on the local dark sector. The limitation of this method similarly comes down to the sensitivity of the Doppler link between Earth and the orbiting spacecraft.

The goal of this paper is to determine the required noise suppression  in order to place stringent constraints on the local dark sector in light of prospective ice giant missions, as well as providing a proof of concept analysis of the trajectory reconstruction method. While our main focus is DM, we also include a limited analysis of several more exotic MG alternatives. The structure of this work is as follows: In Section \ref{sec:plan}, we briefly outline the mock ice giant mission timeline and trajectory and describe how Doppler ranging can determine the influence of the dark sector on the motion of planets.
In Section \ref{sec:ITP} we use Monte Carlo–Markov Chain methods to infer the local DM density from simulated spacecraft trajectories.
In Section \ref{sec:ephe} we follow the path laid in \cite{sereno_dm}, and determine what bounds a prospective ice giant mission can  place on DM and MG theories through extra-precession measurements.
We discuss our findings in Section \ref{sec:disc} and present our concluding remarks in Section \ref{sec:conclusion}. This work is a companion piece to \citet{soyuer2021} (hereafter \citetalias{soyuer2021}),
in which the potential to use the same missions as gravitational wave detectors is investigated.

\section{A Ranging mission to the ice giants}
\label{sec:plan}
\subsection{Mission plan}
\begin{figure}
    \centering
    \includegraphics[width = \columnwidth]{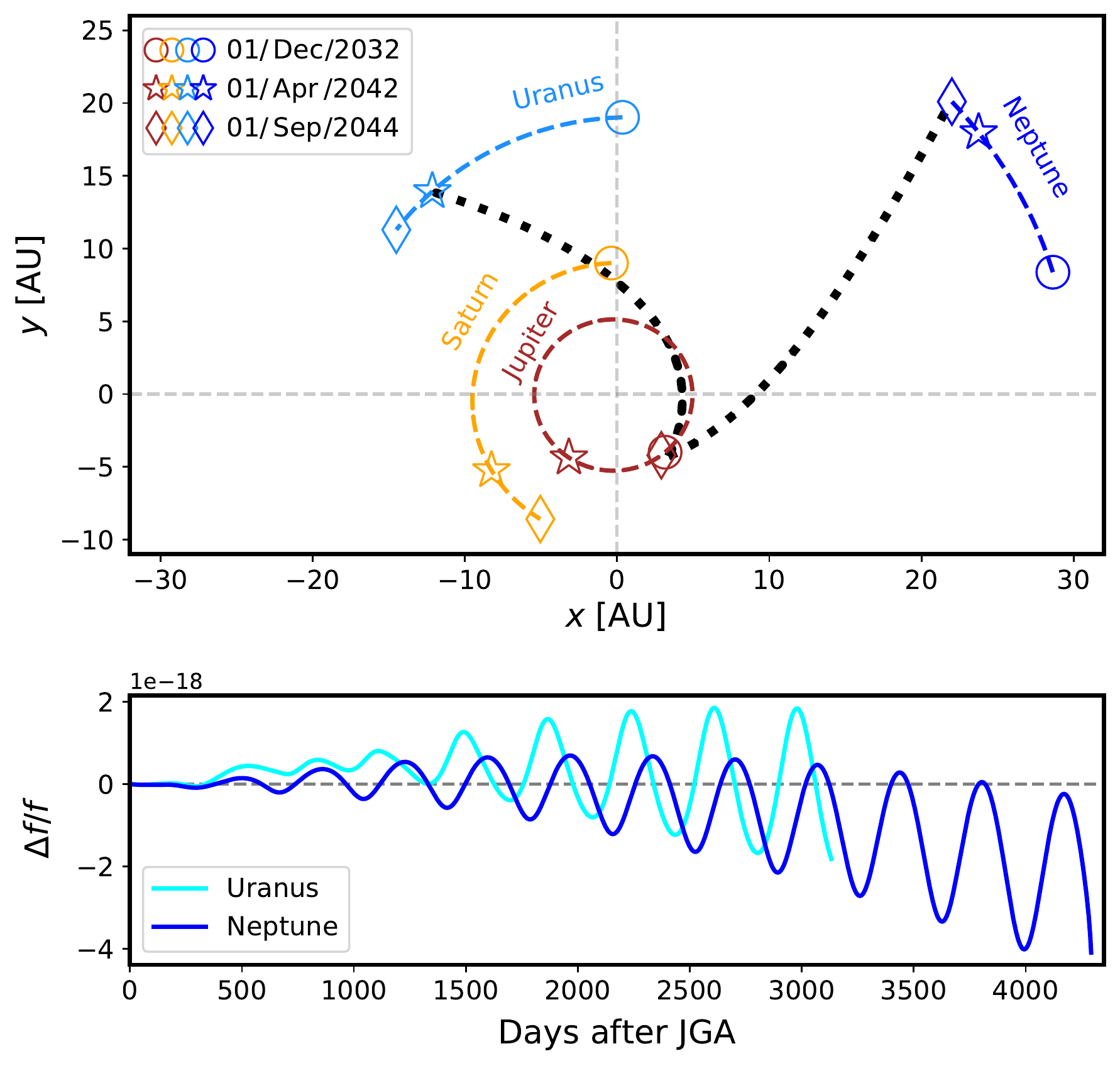}
    \caption{\textbf{Mission plan and Doppler-tracking signal. Top panel: Trajectory of the two spacecraft after the Jupiter swing-by}, where different shapes show the positions of the outer planets at various timestamps in the ICRF. The colourful dashed lines show the orbits of the planets and the black dotted lines show the trajectories of the satellites over a period of approximately ten years. The orbits of the planets are taken from the JPL HORIZONS database using the ASTROQUERY tool by \citet{astroquery}, while the spacecraft trajectories are integrated.
    \textbf{Bottom panel:  Noiseless realisation of the change in the two way frequency fluctuation of  missions to Uranus and Neptune} caused by a constant local dark matter density of  $4.6 \times 10^{-22}$ kg/m$^{3}$. The yearly oscillatory behavior is due to the  revolution of the Earth around the Sun.}
    \label{fig:traj}
\end{figure}

There are many proposed mission outlines for prospective ice giant missions.
A notable one, which is also considered in \citetalias{soyuer2021}, consists of a spacecraft separating into two just before engaging in a Jupiter gravity assist (JGA). Both spacecraft subsequently travel toward their destined planets as shown in Figure \ref{fig:traj}. The mission outline is given as:
\begin{itemize}
    \item \textbf{Feb. 2031:} Space Launch System departure from Earth.
    \item \textbf{Dec. 2032:} Separation of the spacecraft and subsequent JGA.
    \item \textbf{Apr. 2042:} Arrival of the first spacecraft  at Uranus.
    \item \textbf{Sep. 2044:} Arrival of the second spacecraft at Neptune.
\end{itemize}
In this paper, we integrate the trajectories such that the spacecraft arrive at the ice giants at the given timestamps. In the coordinate system depicted in Figure \ref{fig:traj}, the spacecraft are initialised just outside Jupiter's sphere of influence opposite to the Sun, with a velocity of (8.38, 16.75) km/s and (17.65, 7.50) km/s respectively in the International Celestial Reference Frame (ICRF). These initial conditions are chosen in accordance with the proposed mission timestamps\footnote{Provided in {\url{https://github.com/ice-giants/papers/raw/master/presentation/IGs2020_missiondesign_elliott.pdf}}}, and result in the spacecraft reaching the ice giants with velocities of approximately 20 km/s, similar to the Voyager 1 and 2 mission fly-bys.

\subsection{Doppler tracking}
Interplanetary trajectories are monitored by recording Doppler shifts in the time series of the radio link between Earth and the spacecraft.
The precision of this measurement is limited by the noise on the two-way frequency fluctuation $y_2 = \Delta f/f_0$, where  $f_0$ is the carrier frequency of the Doppler link and $\Delta f$ is the discrepancy from $f_0$ measured  on the ground. Therefore, the uncertainty on the measured radial velocity $v_{\rm r}$ is expressed as
\begin{equation}
    \Delta v_{\rm r} = c \Delta y_{2},
\end{equation}
where $c$ is the speed of light and $\Delta y_2$ is the noise of the frequency fluctuation. For the purposes of this paper, this quantity is crucial since it will determine the possible bounds on the local DM content and the effects of MG theories in the solar system. Following \citetalias{soyuer2021}, we take the nominal noise values of the Cassini mission as a baseline reference \citep{cassini92,armstrong}. We express the spread  of the fluctuation as a scaling law with respect to carrier frequency $f_0$, Allan deviation $\sigma_{\rm{A}}$ and two-way light travel time $T_2$:
\begin{equation}
    \Delta y_2 \approx 6\times 10^{-13} \times \frac{f_0^{\rm{Cass}}}{f_0} \frac{\sigma_{\rm{A}}}{\sigma_{\rm A}^{\rm{Cass}}}\sqrt{\frac{T_2}{T_2^{\rm{Cass}}}},
\end{equation}
where $\sigma_{\rm A}^{\rm{Cass}} \approx 3 \times 10^{-15}$ and $T_2^{\rm{Cass}} = 5730$s. This formula suggests that the frequency fluctuation noise can be improved by either reducing the Allan deviation or upgrading the Doppler link to higher frequencies. We discuss both opportunities in more detail in Section \ref{sec:disc}. For the remainder of this paper, we assume for simplicity that the frequency of the link is fixed at the Ka-band (approximately 32 GHz) and express our results solely as a function of the Allan deviation.

\section{Interplanetary transfer phase}
\label{sec:ITP}
As detailed in the previous section, proposed mission plans require  roughly a ten year interplanetary cruise phase after a JGA. In this phase, the spacecraft trajectories will mainly be determined by the configuration of massive bodies in the solar system, as well as the initial conditions when they leave Jupiter's sphere of influence.
The presence of DM affects the trajectory by introducing a small radial acceleration, which in turn reduces the velocity of the spacecraft over many years of interplanetary travel. A perfect ranging system could use this change in velocity to reconstruct the slope and quantity of dark matter around the Sun. 

\subsection{Order of magnitude estimate}

In reality, measuring this velocity difference is extremely difficult, as it lies in the nm/s  for realistic quantities of dark matter. To appreciate the challenge, we present here a short back-of-the-envelope calculation. In the case of a constant background density of dark matter, a spacecraft with mass $m$ exiting Jupiter's sphere of influence will lose additional kinetic energy as it climbs out of the Sun's potential well.
We can estimate the amount of energy by the following formula:
\begin{align}
    \Delta E_{\rm{kin}} \approx G M_{\odot} m \left( \frac{1}{r_{\rm J}}- \frac{1}{r}\right) + \frac{4 \pi}{3}  \rho_{\rm{DM}}Gm\left(r_{\rm J}^2 - r^2 \right),
\end{align}
where $r$ is the spacecraft's distance from the Sun, $r_{\rm J}$ is Jupiter's mean orbital distance and $\rho_{\rm{DM}}$ is the mean density of dark matter around the Sun.
The second term in the expression is purely caused by dark matter, and represents an additional energy loss. This effect translates into a small loss in velocity $\Delta v$, which reads:
\begin{align}
     \Delta v  \approx \frac{8 \pi}{3 v} G \rho_{\rm{DM}} r^2,
\end{align}
where we have assumed that $r \gg r_{\rm{J}}$ and $\Delta v \ll v$.
Considering a mission to one of the ice giants, we expect a change in the  velocity of the order nm:
\begin{align}
    \label{eq:deltav}
    \Delta v \sim 3.2 \left(\frac{\rho_{\rm{DM}}}{10^{-21} \, \frac{\rm{kg}}{\rm{m}^3}}\right)\left(\frac{10 \, \frac{\rm{km}}{\rm{s}} }{v}\right) \, \left[\frac{\rm{nm}}{\rm{s}} \right].
\end{align}
Here, the scaling with 10 km/s  represents typical speeds of interplanetary spacecraft as they approach the ice giants, while the DM density is scaled with expected solar system values from galactic DM halo models. We can compare this change in velocity with the uncertainty of the spacecraft's radial velocity caused by noise in the link. At a distance of approximately 30 AU, the uncertainty reads as follows:
\begin{align}
    \label{eq:deltavlink}
    c \Delta y_2 = 0.2 \frac{\sigma_{\rm{A}}}{\sigma_{\rm A}^{\rm{Cass}}}\, \left[\frac{\rm{mm}}{\rm{s}} \right].
\end{align}
Comparing equations (\ref{eq:deltav}) and (\ref{eq:deltavlink}) suggests that an improvement of at least 5 orders of magnitudes in Doppler tracking precision is required to detect the influence of DM on the trajectory of a spacecraft, and that a mission could only bound the density of DM to:
\begin{align}
    \label{eq:badscaling}
    \rho_{\rm{DM}} \lesssim 3.2\times 10^{-16}\frac{\sigma_{\rm{A}}}{\sigma_{\rm A}^{\rm{Cass}}} \left(\frac{v}{10 \, \frac{\rm{km}}{\rm{s}} }\right)\,\left[ \frac{\rm{kg}}{ \rm{m}^3}\right].
\end{align}

Several order of magnitudes in sensitivity could be gained by choosing trajectories which reach separations of 30 AU with lower velocities. This is however not a reasonable option, since it would require for the mission duration to be extended by a factor approximately equal to the gain in sensitivity. However, the above calculations do not take into account that DM would produce a \textit{systematic} deceleration, while noise is more likely to be randomly distributed. Indeed, integrating the change in velocity over a mission duration of 10 years yields trajectory deviations  of the order of meters:
\begin{align}
    10 \, \rm{yr} \times \Delta v \sim 5.0\times 10^{-1} \left(\frac{\rho_{\rm{DM}}}{10^{-21} \, \frac{\rm{kg}}{\rm{m}^3}}\right)\left(\frac{10 \, \frac{\rm{km}}{\rm{s}} }{v}\right) \, \left[\rm{m} \right].
\end{align}
Positional estimates with a precision of tens of meters at 50 AU have already been achieved by the New Horizons mission, with technologies that were envisioned almost two decades ago \citep[see e.g.][]{newhor08,newhor13}.
This is suggestive that the improvements required by a prospective ice giant mission to detect the influence of DM on the spacecraft trajectory are achievable, provided that they become one of the priorities during mission development.

\subsection{Mock solar system and monte carlo--Markov chain}
We want to provide a more precise estimate on required ranging capacity improvement that would assure a detection of the local DM content. For this purpose we developed a numerical procedure based on a Monte Carlo--Markov Chain (MCMC). The procedure consists in reconstructing the influence of DM in the Doppler signal of thousands of simulated ice giant missions. The virtual spacecraft are equipped with ranging systems with differing Allan deviations, from Cassini-era values of $\sigma^{\rm Cass}_{\rm A} \sim 3 \times 10^{-15}$ to improvements of order $10^3$. The spacecraft are tracked from the moment they exit Jupiter's sphere of influence to when they reach Uranus and Neptune. They are subject to the force of gravity, and their trajectory is determined by a symplectic integrator, assuring the conservation of energy.
We consider a simplified solar system model consisting of only the outer planets, the Sun and finally a constant density of DM. The positions of the outer planets are not integrated, but rather sampled at the required time-steps from the JPL HORIZONS database.
In total, we therefore consider six free parameters in our MCMC chains.

\begin{figure*}
    \centering
    \includegraphics[width = \textwidth]{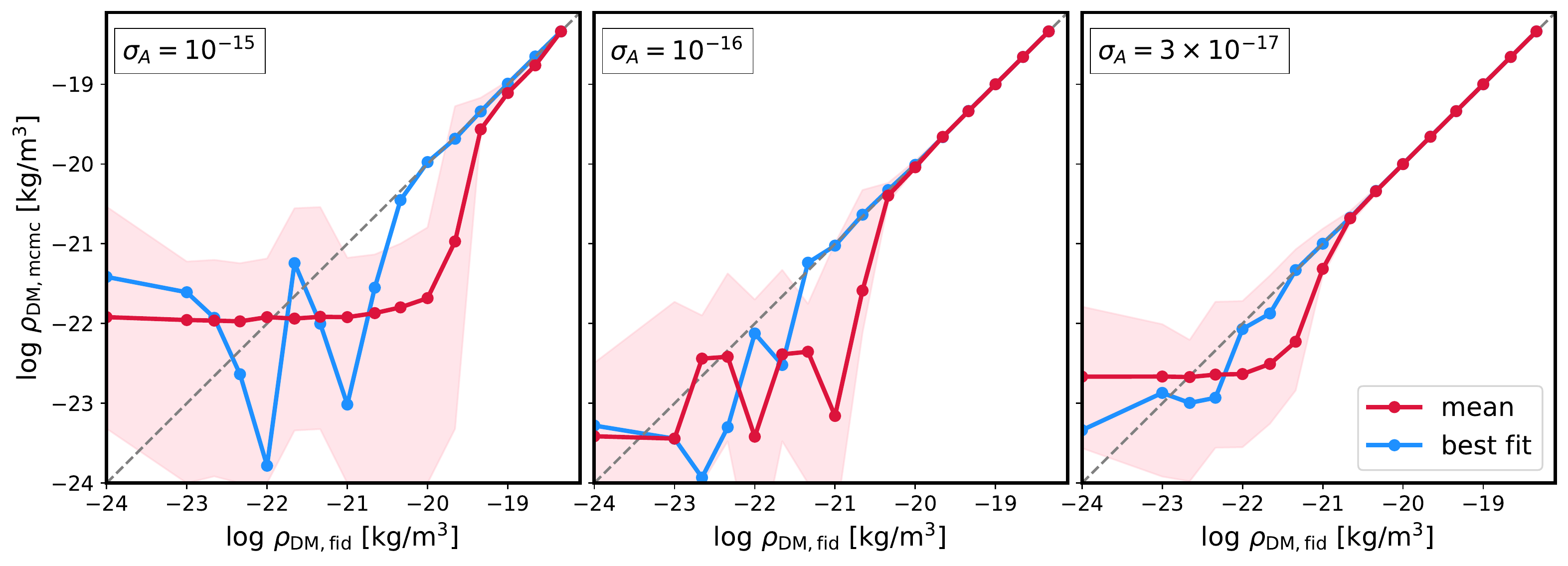}
    \caption{\textbf{Reconstructed values of DM density as a function of the true density of DM in the solar system, for an improvement of 3 (left), 30 (centre) and 100 (right) times the Cassini-era Allan deviation.} The blue lines show the most probable value given by the marginalised posteriors, while the red lines show the mean value. The $1\sigma$ uncertainty contour is shown by the shaded area. As expected, missions with an improved Allan deviation are sensitive to smaller quantities of DM. As a criterion for detectability, we check whether both the mean and the most probable posterior values converge on the true DM density. This criterion defines a mission sensitivity, which is approximately given by the scaling $\rho_{\rm{MS}} \sim 9 \times 10^{-20} \, (\sigma_{\rm A}/\sigma_{\rm A}^{\rm{Cass}})$ kg/m$^3$.   }
    \label{fig:rhoDM_fid_vs_mcmc}
\end{figure*}

To perform the MCMC analysis, we use the sampler \texttt{emcee} \citep{emcee} based on the 'stretch move' method \citep{camcos}.  We run the chains on the Eiger supercomputer, which is a part of Swiss National Supercomputing Centre. Each of the 128 parallel MCMC walkers for the standard gravitational parameters of the Sun and four giant planets is initialised under uniform probability distributions on a full domain centred on the values taken from \citep[][]{park21}, with the width of 20 standard deviations. DM densities can vary from $10^{-24}$ to $10^{-18}$~kg/m$^3$ with the expected galactic value being close to the middle of the prior domain. In most cases, we initialise the DM density parameter similarly to the other ones, using a uniform prior over the full domain. However to speed up the convergence in some specific setups, we use Gaussian priors centered on the fiducial (ground truth) value with the standard deviation of 5\% of the considered mean. To asses the convergence of our chains we use the Gelman-Rubin criterion \citep[][]{GelmanRubin1992}, with $R_c \leq 1.1$. To define an MCMC likelihood, we model the trajectories of both spacecraft and the resulting Doppler link signal for a given configuration in the 6-dimensional domain of MCMC parameters. We use the residuals of the Doppler link time-series between  solar systems with a randomly sampled and the fiducial amount of DM as the signal for the MCMC analysis. The log-likelihood is described by:

\begin{equation}
     \log \mathcal{L (\theta)} = -\frac{1}{2}\left( \sigma^c(\theta)  - \sigma^c_{\rm GT}   \right) \Sigma^{-1} \left( \sigma^c(\theta) - \sigma^c_{\rm GT} \right)^T,
\end{equation}
where $\sigma^c(\theta) = (\sigma_{\rm U} (\theta),\sigma_{\rm N}(\theta))$ is a concatenated Doppler link signal from both spacecraft for a given parameter space configuration $\theta$ and the subscript GT denotes the signal modelled for fiducial values for the parameters. The signal vector is composed by subsequent measurements made during the travel of spacecraft from Jupiter to Uranus and Neptune, with a temporal resolution of one day. An example of the Doppler shift signal can be seen in Figure \ref{fig:traj}, in which a virtual mission is launched in a solar system with a DM density of $\rho_{\rm DM} = 4.6\times 10^{-22}$ kg/m$^3$.

In our study, we consider the Allan deviation to be descriptive of the dominant noise sources on the Doppler link signal. We pollute the signal at every time-step with random noise, sampled from a Gaussian distribution with a standard deviation given by the Allan deviation of the link. Furthermore, we assume that measurements from both spacecraft are independent and have the same Allan deviation, and therefore define $\Sigma := {\rm diag} (\sigma_{\rm A}^2)$.
We optimise the run-time of the chain by employing OMP parallelism and run each walker on a separate thread using the \texttt{multiprocess} package \citep{multi}. Finally, to overcome the floating point precision limit due to the very small DM densities probed, we use the \texttt{mpmath} \citep{mpmath} python library to implement floating-point arithmetics with arbitrary precision, which we require up to 25 digits. The caveats and limitations of our MCMC approach are discussed in Section \ref{sec:disc}.

\subsection{Results}
In Figure \ref{fig:rhoDM_fid_vs_mcmc} we show the reconstructed values for the local DM density as a function of the true density in the solar system. We probe three values of Allan deviation, which represent an improvement of 3, 30 and 100  times  from Cassini-era technology. Missions with a reduced Allan deviation are more sensitive to the presence of DM, as expected. As a criterion for the detection of a given quantity of DM, we require that both the most probable and the mean reconstructed DM density converge on the true value. Then, the sensitivity of a prospective mission to the local DM density follows a similar scaling to equation (\ref{eq:deltavlink}), improving linearly with Allan deviation. We can estimate an approximate scaling relation, which reads:
\begin{align}
    \label{eq:dmgoodscaling}
    \rho_{\rm{MS}} \sim  9\times 10^{-20}\frac{\sigma_{\rm{A}}}{\sigma_{\rm A}^{\rm{Cass}}} \,\left[ \frac{\rm{kg}}{ \rm{m}^3}\right],
\end{align}
where the subscript MS stands for 'mission sensitivity'. Note that the prefactor $9\times 10^{-20}$ is specific to the initial conditions specified in Section \ref{sec:plan}, and is bound to change if the trajectory of the prospective mission is significantly different than the one we have assumed. Moreover, it is limited to the simplified solar system detailed in the previous section. Nevertheless, it is suggestive of the fact that an improvement of between two and three orders of magnitude is required to confidently achieve expected galactic DM densities of $\sim 4.6 \times 10^{-22}$ kg/m$^3$.
\begin{figure}
    \centering
    \includegraphics[width = 1\columnwidth]{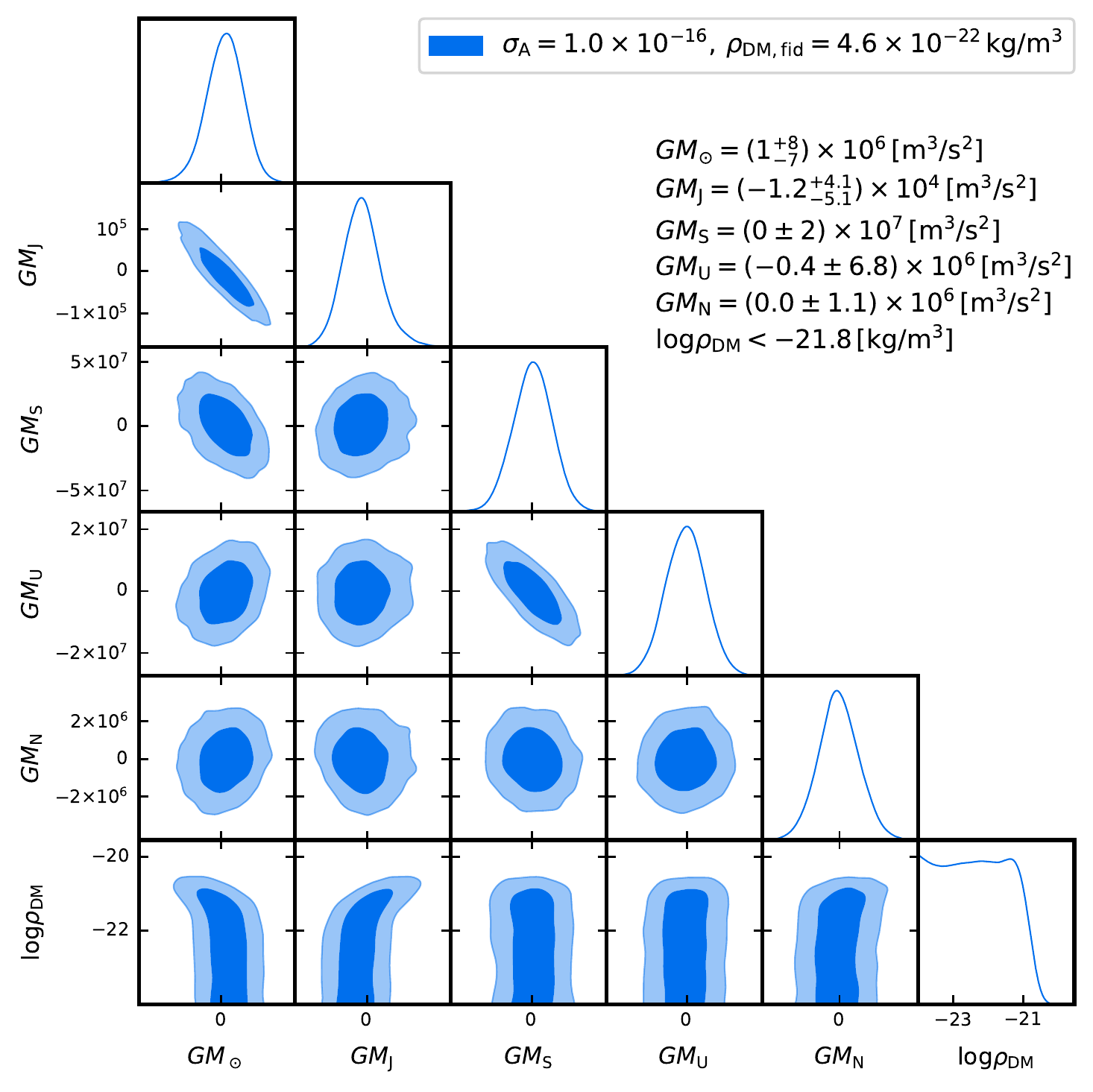}
    \caption{\textbf{Posteriors of the MCMC parameters for a specific choice of DM density and Allan deviation.} In this example, the mission provides an upper bound of $\rho_{\rm{DM}} \lesssim 10^{-21}$ kg/m$^3$, but is not sensitive enough for a clear detection. The standard gravitational parameters of our simplified solar system model are measured with extreme precision. While this is likely due to the simplicity of our Solar system model, it is suggestive of the potential of a future ranging mission to measure planetary and solar masses.}
    \label{fig:corner_mcmc}
\end{figure}

\begin{figure}
    \centering
    \includegraphics[width = 1\columnwidth]{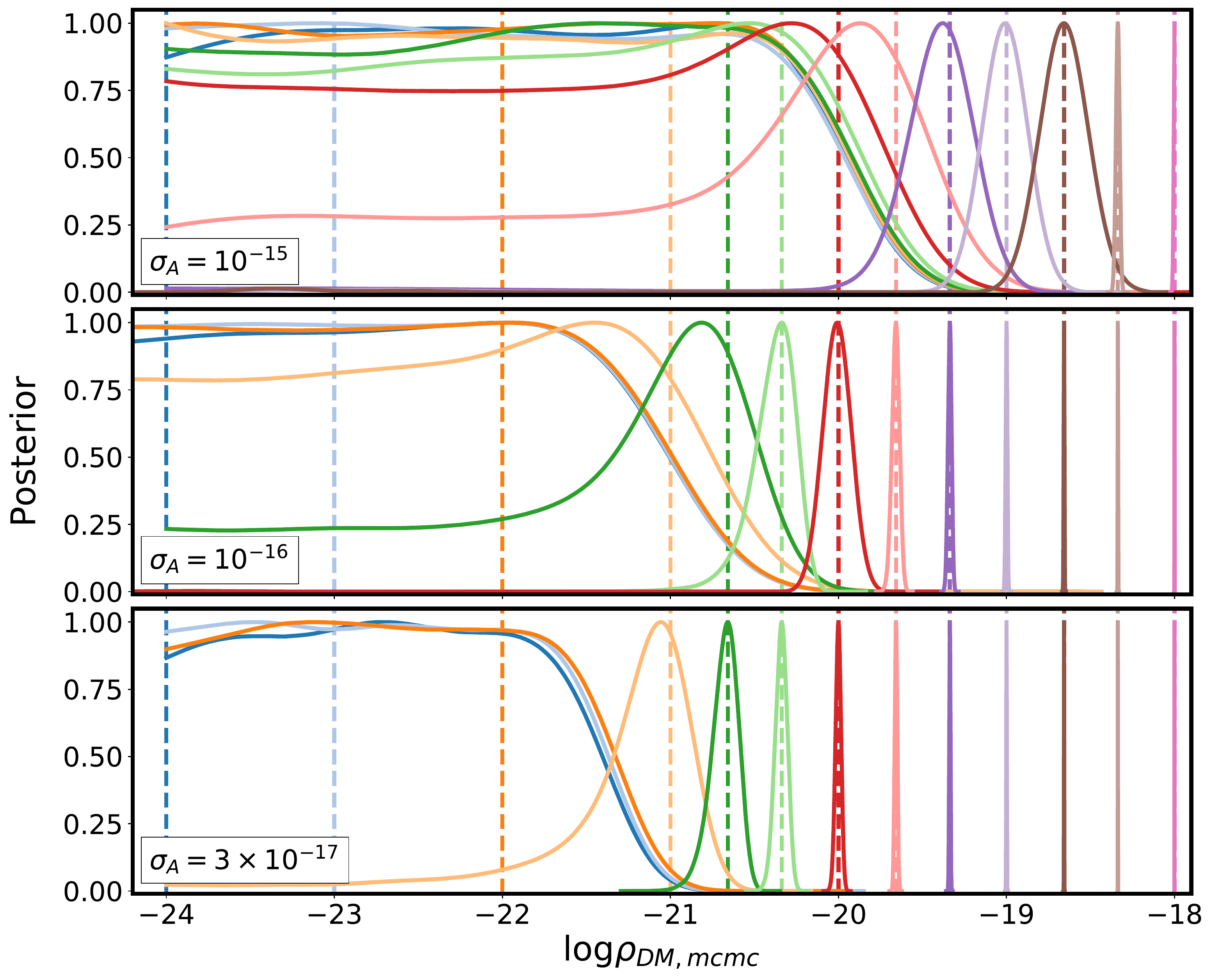}
    \caption{\textbf{Posteriors of the DM density for different fiducial values and mission sensitivities}. The fiducial values are represented by vertical dashed lines of different colors. The posteriors belonging to the same run share the colours. Note that, for larger DM densities or lower Allan deviations the posteriors smoothly transition from a step function to a Gaussian. Our criterion for producing equation (\ref{eq:dmgoodscaling}) is the presence of a clear Gaussian centered on the true value.}
    \label{fig:rho_posterior}
\end{figure}
\begin{figure}
    \centering
    \includegraphics[width = \columnwidth]{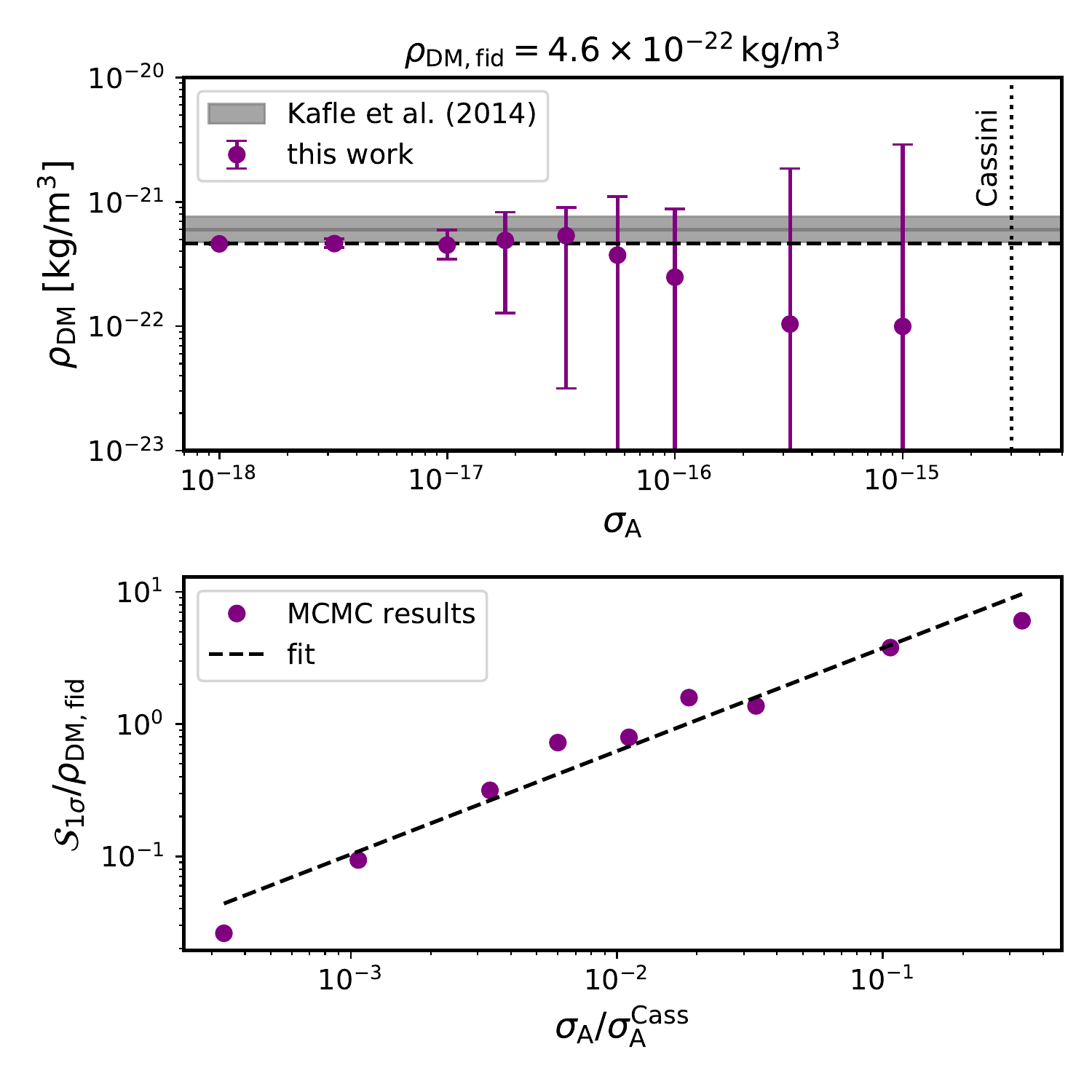}
    \caption{\textbf{MCMC results for the expected galactic density. Top panel: Constraints for local dark matter density as a function of Allan deviation.} The dots signify the most probable value of the posterior distribution, while the errorbars show the $1\sigma$ uncertainty contours. Note how the upper and lower contours become symmetric with improving Allan deviation. We also show current galactic-based constraints given in \cite{kafle14} as a black dot for reference. \textbf{Bottom panel: Power-law fit of the relative 1$\mathbf{\sigma}$ upper bound uncertainty.} The best fit parameters for the form $A (\sigma_{\rm A}/ \sigma_{\rm A}^{\rm{Cass}})^n$ are $A = 22.7$ and $n = 0.78$ (with a fit goodness of 0.90), which we round to 23 and 0.8 respectively. }
    \label{fig:errors_vs_sigma}
\end{figure}

A more detailed view of the MCMC procedure can be observed in Figures \ref{fig:corner_mcmc} and \ref{fig:rho_posterior}. In the former, we show a realisation of a converged MCMC corner plot for $\sigma_{\rm A} = 10 ^{-16}$ and $\rho_{\rm DM} = 4.6 \times 10^{-22}$ kg/m$^3$. The two values are chosen for visualisation purposes, so that the posterior on the DM density does not peak around the true value, but rather only provides an upper bound. Interestingly, we find that the posteriors on the standard gravitational parameters of the outer planets and the Sun are extremely well converged, providing constraints that are several orders of magnitude more precise than current estimates \citep[see e.g.][along with the JPL Horizons database]{std92}. While this is likely a result of our simplistic solar system model, it is suggestive that prospective ice giants missions could provide extremely precise constraints on gravitational dynamics in the solar system. 

In Figure \ref{fig:rho_posterior} we show several realisations of the marginalised DM density posterior for different Allan deviations and different DM density values. Converging the posterior on a specific value for the DM density is only possible if the Allan deviation of the mission is low enough, following the behaviour described by equation (\ref{eq:dmgoodscaling}). The emerging long, flat probability tails for insufficiently precise missions introduce a significant difference between the most likely value and the mean value of the posteriors.

Finally in Figure \ref{fig:errors_vs_sigma}, we specifically focus on the scenario in which the local DM density is equal to the expected galactic mean. As the Allan deviation of the mission improves, the upper and lower $1\sigma$ uncertainties of the reconstructed value become more symmetric, signifying how the posterior approaches a Gaussian distribution. We find that the $1\sigma$ upper bound $\mathcal{S}_{1\sigma}$ can be fit by 
\begin{align}
    \mathcal{S}_{1\sigma}  \approx 1.04 \times 10^{-20} \left(\frac{\sigma_{\rm A}}{\sigma_{\rm{A}}^{\rm Cass}} \right)^{0.78} \, \left[ \frac{\rm{kg}}{\rm{m}^3} \right].
\end{align}

Targeting an Allan deviation of $\sigma_{\rm A}$ $\sim 10^{-17}$ would assure the detection of DM with uncertainties of roughly $20\%$ of the total value, comparable to those of  recent galactic estimates \citep[e.g.][]{kafle14}. While technologically ambitious, this is an extremely worthwhile goal since similar improvements would also likely lead to the detection of at least several sources of gravitational waves during the cruise phase of the mission (as shown in \citetalias{soyuer2021}).



\section{\textit{In-situ} measurements}
\label{sec:ephe}
Depending on the science objectives, prospective  missions to the ice giants  are likely to be orbiters  around Uranus and Neptune.
A Doppler link to an orbiter presents a unique opportunity for measuring the precession rate of planets, which will be affected by the presence of DM and by modifications to the inverse square law \citep{sereno_dm}. Typically, the influence of DM or MG has been searched in a quantity called "extra-precession". It is defined as the observed rate of phase shift of an orbit's true anomaly with respect to the most precise solar system simulation available.
Extra-precession measurements of Uranus and Neptune are currently limited to astrometric observations. Currently, the uncertainties of both measurements and simulations are of  comparable order \citep[see e.g.][]{pitjeva_2013,Willgravitoni}, and therefore inconclusive on the matter of DM and MG. Rather, what is commonly done in the literature is to \textit{assume} that the current uncertainties are equal, and uniquely caused by the presence of unmodelled DM or MG theories. While undoubtedly an over-simplification, this can be used to provide an order of magnitude constraint on the importance of the dark sector within the solar system. In what follows, we apply this procedure to several classes of MG theories and estimate the bound that a prospective ice giant missions can place on them. We discuss the limitations of this approach in more detail in Section \ref{sec:disc}.

\subsection{Perihelion precession of Uranus and Neptune}
In the following, we assume for simplicity that the ranging satellites are placed at the centre of mass of the Uranus and Neptune.
Given the fact that Doppler tracking only provides a radial measurement of velocity, we have to perform a few extra steps to constrain the extra-precession rate $\delta\dot{\varphi}_{\rm p}$ of Uranus and Neptune. We use the JPL HORIZONS database to provide a set of functions $R_{i}(t)$, which return the radial distances between the Earth and the ranging satellites. We can then re-parametrise $R_i(t)$ with the orbital phases $\phi_i(t)$ of the ice giants with respect to an arbitrary axis originating from the Sun. Given a radial measurement $r_i$ of the distance between the Earth and the planet in question, one can reconstruct the phase of the planet by inverting the function $R_i(\phi_i)=r_i$, which can be compared with the expected value.

The limiting factor in this process is the precision with which a ranging spacecraft can measure the exact distance between the Earth and the planet. A radial measurement of distance can be performed by determining the light travel time between the Earth and the target satellite. In principle, such a measurement is only limited by the Allan deviation of ground based clocks. In the case of a ranging spacecraft however, this precision is degraded due to various factors such as the mechanical noise of the ground based antenna, interplanetary plasma scintillation and tropospheric noise \citep{armstrong}.
In other words, it is once again the noise on the Doppler link;  $\Delta r= c T_2 \Delta y_2$ that determines the precision of a distance measurement.

\begin{figure}
    \centering
    \includegraphics[width = 0.5 \textwidth]{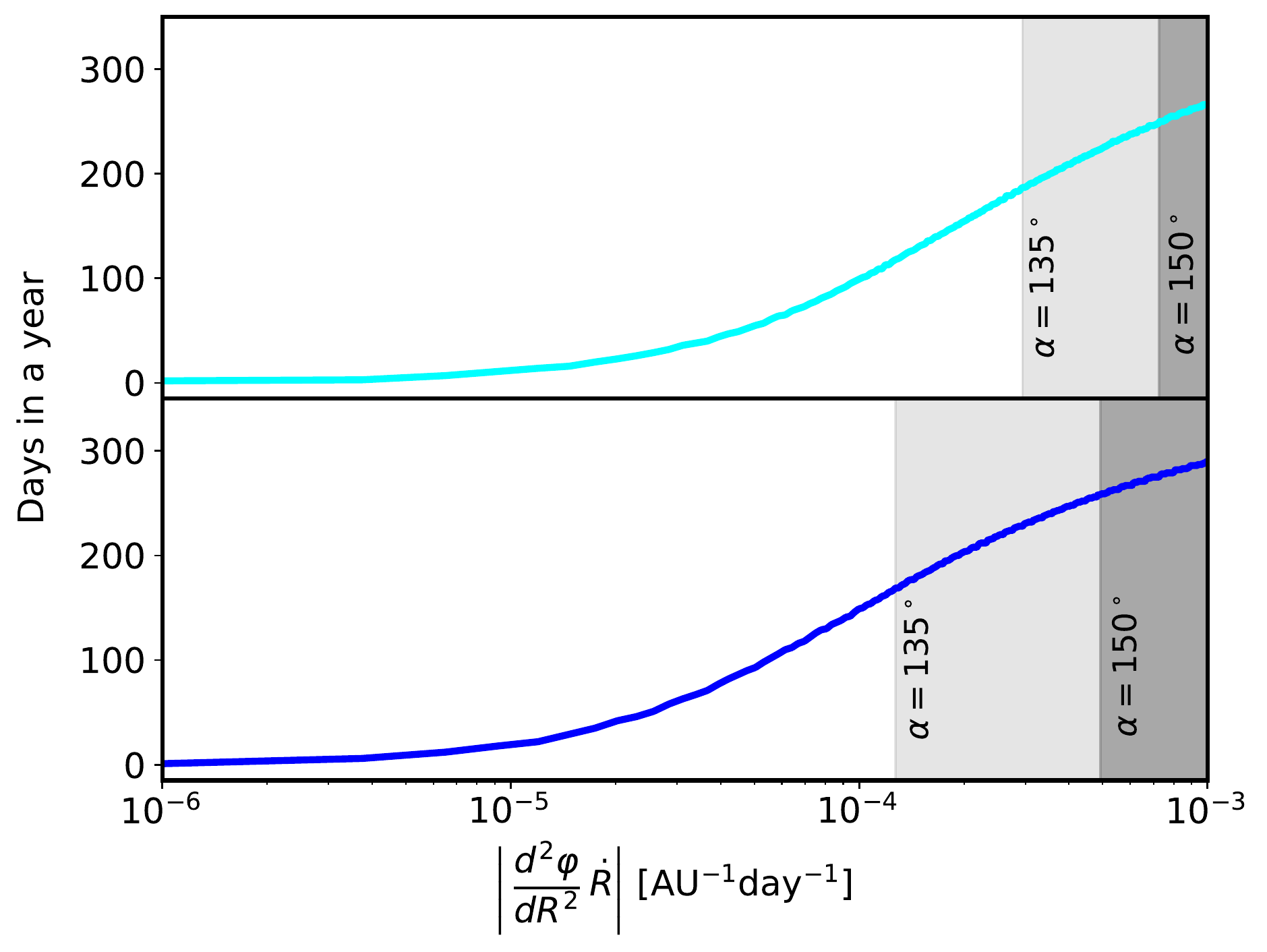}
    \caption{\textbf{Total days in a year for which the prefactor of the uncertainty $\Delta R$ in equation (\ref{eq:phi_rr}) is below a certain level} (cyan for Uranus and blue for Neptune). Since the prefactor is directly related to the arrangement of the Sun, Earth and the spacecraft, the grey shaded areas correspond to the $(d^2 \varphi/dr^2) \dot{R} $ values  where the Sun--Earth--spacecraft angle  $\alpha$ is above $135^{\circ}$ and $150^{\circ}$, respectively.}
    \label{fig:phi_rr}
\end{figure}

For the remainder of this section, we define the extra-precession $\delta \dot{\varphi}_{\rm p}$ as the small additional angular velocity of each ice giant due to DM or MG. Given a radial measurement uncertainty of $\Delta r$ we can find the uncertainty in the reconstructed angular velocity $\omega$ using Gaussian propagation:
\begin{align}
    \Delta \omega = \left(\frac{d^2 \varphi}{dR^2}\dot{R}\right) \Delta r.
    \label{eq:phi_rr}
\end{align}
The prefactor $(d^2 \varphi/dr^2) \dot{R} $ is calculated numerically from our simplified solar system model based on the JPL HORIZONS database, and can vary by several orders of magnitude for different configurations of the Sun--Earth--spacecraft system. Since precession measurements do not require constant monitoring of the spacecraft, they could in principle be done at orbital phases where the prefactor becomes minimal. As mentioned in \citetalias{soyuer2021} and \citet{armstrong} however, the optimal Sun--Earth--spacecraft angle for minimising plasma scintillation noise due to interplanetary dust is > 150$^\circ$. This severely limits the optimal window for a measurement of the precession rate. Figure \ref{fig:phi_rr} shows how many days in a year $(d^2 \varphi/dr^2) \dot{R} $ is lower than a certain value, as well as the values that correspond to  certain convenient configurations for the purposes of minimizing plasma scintillation noise. For the remainder of this section, we take the pre-factors $(d^2 \varphi/dr^2) \dot{R} $ to be
\begin{equation}
    \left(\frac{d^2 \varphi}{dR^2}\dot{R}\right)_{U, (N)} = 3  \times 10^{-4}, \,(1.5  \times 10^{-4}) \,\,\,\left[ \rm{AU}^{-1} \rm{day}^{-1} \right].
\end{equation}
This choice corresponds to values shown in Figure \ref{fig:phi_rr}, where the Sun--Earth--spacecraft  angle $\alpha$ is above 135$^\circ$, representing a compromise between the plasma scintillation noise being sufficiently low and the pre-factor being as small as possible. As can also be seen in Figure \ref{fig:phi_rr}, a more conservative choice of $\alpha > 150^{\circ}$ would worsen the constraints by a factor $\sim$3 to $\sim$5.

Now that we have an estimate of how precisely Doppler ranging can reconstruct the precession rate of the outer planets, we compare it with the expected amount of extra-precession caused by DM or MG. From celestial mechanics we can calculate how a small additional radial acceleration $\delta \mathcal{A}_{\rm R}$ affects the motion of a planet. Crucial for our application is the change in angular velocity, which is given by the following formula \citep[see e.g.][]{sereno_dm,sereno_de}:
\begin{equation}
\langle\delta\dot{\varphi}_{\mathrm{p}}\rangle=\frac{\sqrt{1-e^2}}{n a} \delta \mathcal{A}_{\mathrm{R}},
\end{equation}
where $a$ is the semi-major axis, $e$ the eccentricity and $n = \sqrt{GM/a^3}$ for the planet in question.
By plugging in the appropriate perturbations  $\delta \mathcal{A}_{\mathrm{R}}$ associated with DM and MG theories into the above equation, and comparing the outcome with equation (\ref{eq:phi_rr}), one can estimate the constraints that can be placed on both type of theories.
\subsection{Dark matter}
The presence of DM produces an additional radial acceleration proportional to the total enclosed mass within the sphere defined by a planet's orbit. In the case of a constant density profile it reads:
\begin{align}
    \delta \mathcal{A}_{\rm R} = -\frac{4 \pi G \rho_{\rm{DM}}}{3}r.
\end{align}
This acceleration causes an extra-precession, given here both in terms of dark matter density and total enclosed mass $\mathcal{M}_{\rm{DM}}$:
\begin{align}
\label{eq:dmep}
\langle\delta\dot{\varphi}_{\mathrm{p}}\rangle=\frac{2 \pi G \rho_{\rm{DM}}}{n a} \approx 3 \frac{ G \mathcal{M}_{\rm{DM}}}{2 n a^4},
\end{align}
where we assume simplicity that planetary orbits are circular. By comparing equations (\ref{eq:phi_rr}) and (\ref{eq:dmep}), we can construct scaling relations for the density and total mass of DM for which prospective ice giant missions would be sensitive to:
\begin{align}
    \left(\rho_{\rm{DM}}\right)_{U,(N)} &\leq 2.9 \times 10^{-18}, (1.4 \times 10^{-18}) \frac{\sigma_{\rm A}}{\sigma_{\rm A}^{\rm{Cass}}} \, \left[\frac{\rm{kg}}{\rm{m^3}} \right].
\end{align}
This method yields constraints that approximately one order of magnitude worse than the ones produced by the trajectory reconstruction method during the interplanetary phase (see equation (\ref{eq:dmgoodscaling})). If orbiters were indeed included in a prospective mission, they could provide a competitive independent constraint on the DM content in the solar system.

Bounds on the DM density can be used to improve solar system based constraints on galactic DM halo models, specifically the widely accepted Navarro-Frenk-White (NFW) profile \citep{NFW96}. Within the NFW framework, the DM density  can be described by two parameters, a characteristic density $\rho_{\rm{NFW}}$ and and a scale radius $R_{\rm{NFW}}$:
\begin{align}   
    \label{eq:nfw}
    \rho(r)= \rho_{\rm{NFW}} \frac{R_{\rm{NFW}}}{r\left(1+ {r/R_{\rm{NFW}}} \right)^2}.
\end{align}

\begin{figure}
    \centering
    \includegraphics[scale=0.7]{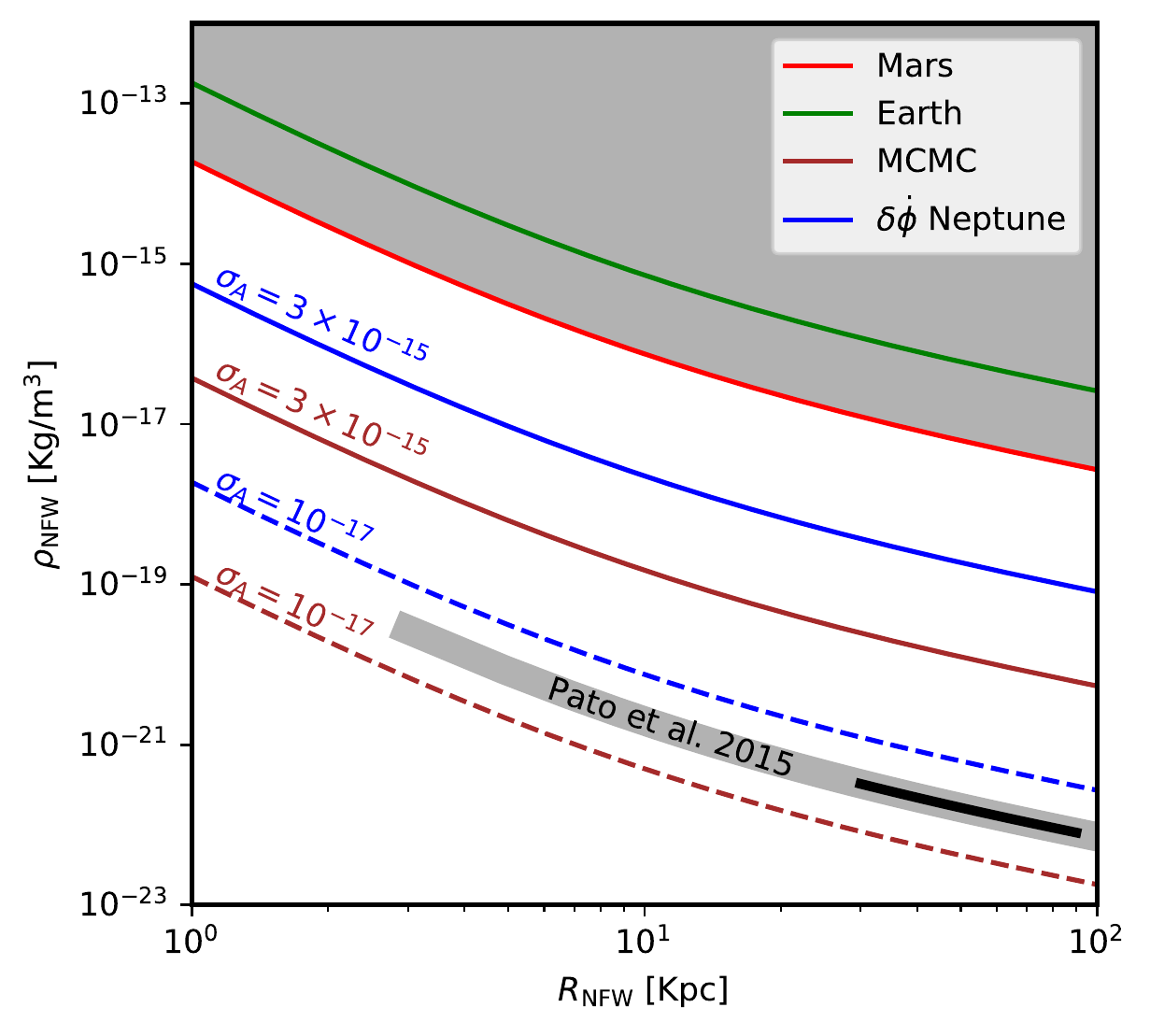}
    \caption{\textbf{2$\sigma$ constraints that can be placed by a prospective ice giant mission, on the two NFW profile free parameters (see equation (\ref{eq:nfw})).} The grey area above the green and red line corresponds to the parameter space that is currently ruled out by local observations of the Earth's (green) and Mars' (red) orbital precession. The solid blue and brown lines denote the constrains that can be placed by a future mission to the ice giants without any improvement from Cassini-era Allan deviation, by using either the orbital reconstruction method (brown) or extra precession measurements (blue). The dashed lines denote the constraints that would be achievable with a target Allan deviation of $\sigma_{\rm A}=10^{-17}$, which would be competitive with current galactic based constraints \citep[e.g.][ shown in the plot at 1$\sigma$ in black and  $2\sigma$ in grey]{pato2015}.}
    \label{fig:nfw}
\end{figure}

Figure \ref{fig:nfw} shows the possible constraints on the characteristic density and the scale radius of the Milky Way's DM halo by a prospective mission to the ice giants. The mission is guaranteed to surpass current extra-precession based constraints by several orders of magnitudes, and can become competitive with galactic based constraints with enough improvement in the Allan deviation.

\subsection{Modified gravity}
Modified gravity theories propose an effective modification to the inverse square law in regimes where the acceleration is low. The interpolation between different regimes is done by  the function $\mu(x)$, where the dimensionless parameter $x$ denotes an acceleration, scaled by a free parameter $a_0$, whose best fit value currently stands at $a_0 = 1.2 \times 10^{-10}$ m/s$^2$. While many interpolation functions have already been disproved by galactic and local constraints \citep[see e.g.][]{mond12,Mond13,Mond18}, solar system observations have yet to rule out the original form proposed by \citet{milgrom83}:
\begin{align}
    \mu(x) = \frac{x}{\sqrt{1+x^2}}.
\end{align}

\begin{figure}
    \centering
    \includegraphics[scale=0.7]{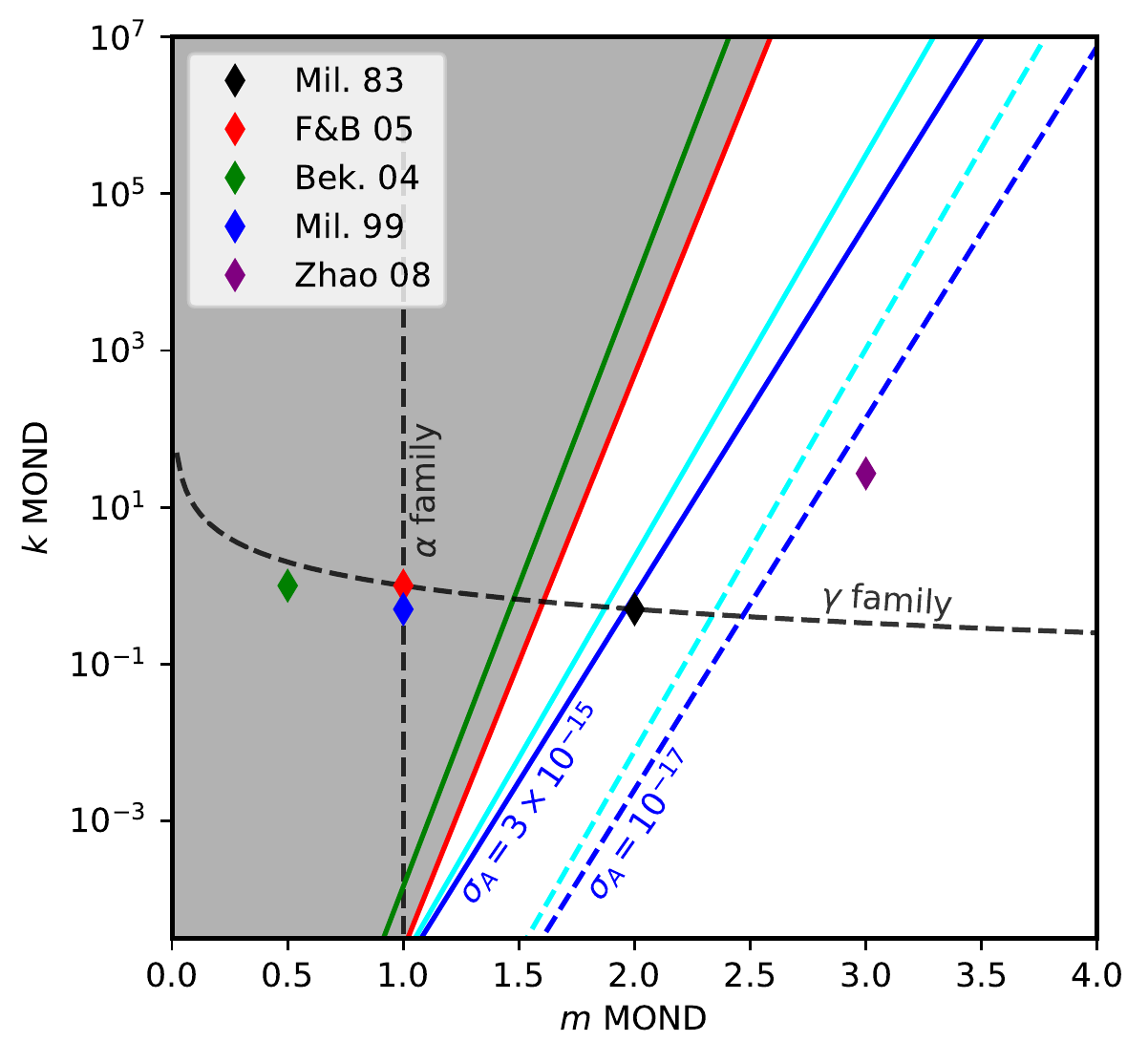}
    \caption{\textbf{2$\sigma$ constraints that can be placed by a prospective ice giant mission on the MOND interpolating function.} The grey area above to the left of the green and red line corresponds to the parameter space that is currently ruled out by local observations of the Earth's (green) and Mars' (red) orbital precession. The solid blue and cyan lines denote the constrains that can be placed by a future mission to the Neptune and Uranus respectively, without any improvement from Cassini-era Allan deviations. The dashed lines denote the constraints that would be achievable with a target Allan deviation of $\sigma_{\rm A}=10^{-17}$. Only a moderate improvement in Allan deviation is required to rule out Milgrom's original interpolating function with solar system based observations. The coloured diamonds stand for several other proposed intorpolating functions \citep{milgrom83,bek05,milgrom99,Zhao08}. The $\alpha$ and $\gamma$ family lines are taken from \cite{MGrev12}.}
    \label{fig:mond}
\end{figure}

Following \citet{sereno_dm},  we analyse a more general class of interpolating functions, which can be approximated by the following expansion $\mu(x) \approx 1+ k x^{-m}$. The effect of the interpolating function on the precession rate of the outer planets can be estimated as:
\begin{align}
    \left< \dot{\omega}_{p}\right>= -  k m \sqrt{2 G M/a^3}\left( \frac{a}{a_{\rm{MOND}}} \right)^{2m}\left[1+ e^2 \left(5m - 2m^2 \right)/4 \right].
\end{align}
This expression can be turned into a constraint in the parameters of the interpolating function itself. As can be seen in Figure \ref{fig:mond}, a prospective mission to the ice giants is guaranteed to improve local constraints on the interpolating functions. An improvement of one order of magnitude in the Allan deviation is required to rule out Milgrom's original proposal with a solar system based measurement. Such an improvement is certainly within reach for a mission that is to be launched in the early 2030s, and is likely to provide conclusive evidence on the DM vs. MG debate.

\subsection{Yukawa-like extensions}
\begin{figure}
    \centering
    \includegraphics[scale=0.7]{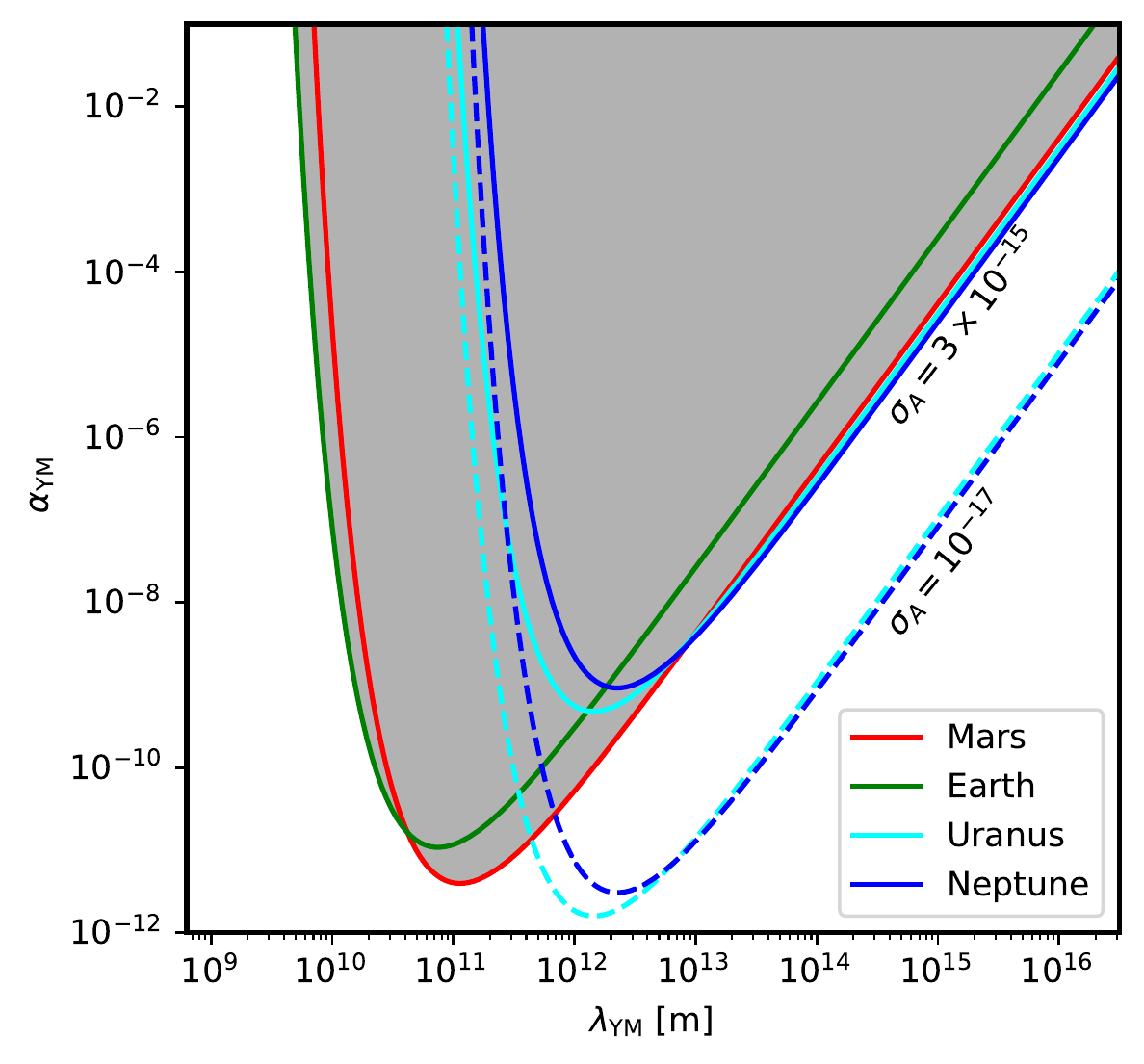}
    \caption{\textbf{2$\sigma$ constraints that can be placed by a prospective ice giant mission on Yukawa-like extensions to the inverse square law.} The grey area above the green and red linen corresponds to the parameter space that is currently ruled out by local observations of the Earth's (green) and Mars' (red) orbital precession. The solid blue and cyan lines denote the constrains that can be placed by a future mission to Neptune and Urnaus respectively, without any improvement from Cassini-era Allan deviations. The dashed lines denote the constraints that would be achievable with a target Allan deviation of $\sigma_{\rm A}=10^{-17}$. Any improvement in Allan deviation is likely to rule out regions of the parameter space that correspond to length scales of the order 30 AU.}
    \label{fig:yukawa}
\end{figure}

Several classes of MG theories can be expressed as a Yukawa-like modification to the gravitational potential, in which the Newtonian gravitational term is enriched with an exponential \citep[see e.g.][]{yuk,yukreview,yuk13,yuk18}:
\begin{align}
    \phi_{\rm{Y}}= \frac{G_{\infty}M}{r} \left[1+ \alpha_{\rm Y} \exp \left(- \frac{r}{\lambda_{\rm Y}} \right) \right],
\end{align}
where $\alpha_{\rm{Y}}$ is a coupling constant and $\lambda_{\rm Y}$ a length scale. For these types of theories, the motion of the outer planets is also modified, producing a small extra precession \citep{sereno_dm}:
\begin{align}
    \left< \dot{\omega}_{p}\right>= \alpha_{\rm Y}\frac{\sqrt{2 G M }}{2\sqrt{a^3}} \exp \left( -\frac{a}{\lambda_{Y} } \right) \left( \frac{a}{\lambda_{Y} }\right)^2\left[1- \frac{e^2}{8} \left(4 - \left( \frac{a}{\lambda_{Y} }\right)^2 \right) \right].
\end{align}
We can transform this equation into a series of constraints on the coupling constant and length scales of Yukawa-like theories. As can be seen in Figure \ref{fig:yukawa}, a prospective mission to the ice giants has the potential to improve such constraints on length scales comparable to the orbits of Uranus and Neptune.

Yukawa-like potentials are intimately connected with the existence of a massive force carrying boson as the mediator of the gravitational force. Solar system based bounds on the graviton mass have been recently updated in \citet{Willgravitoni}. The orbital motion of Mars constrains it below $ \sim 2.5 \times 10^{-23}$ eV/$c^2$. By following the calculations in \citet{Willgravitoni}, we find that a prospective ice giant mission can bound the graviton mass down to:
\begin{align}
    m_{\rm{graviton}} \leq 2.0 \times 10^{-23} \frac{\sigma_{\rm A}}{\sigma_{\rm A}^{\rm{Cass}}} \, \left[\frac{\rm{eV}}{c^2}\right].
\end{align}
Even with no improvements from Cassini era technology, a prospective ice giant mission will provide bounds that are competitive with the current state of the art. However, the linear scaling with Allan deviation suggests the potential to surpass current local and gravitational wave based constraints by several order of magnitudes, though future space based detectors might improve on the latter by even more \citep[see][]{GWgraviton,gwgravitoni}.

\section{Discussion}
\label{sec:disc}
\subsection{Interplanetary transfer phase}
The numerical procedures detailed in Section \ref{sec:ITP} has numerous limitations, which we elaborate on here. First of all, our model of the solar system is far from being complete. We do not consider the gravitational influence of the rocky planets, moons or asteroids. Similarly, more subtle effects such as radiation pressure, the galactic tidal field and modifications due to general relativity are also neglected. Many of these effects are expected to influence spacecraft trajectories more strongly than the presence of DM, which is roughly equivalent to the gravitational pull of $\sim 10^{-13}$ M$_{\odot}$ of enclosed mass. In this sense, our analysis can only be considered a proof of concept, with more expensive numerical simulations being required to strengthen the conclusions. Despite the simplicity however, we believe that our results still provide a reliable estimate.

Our procedure consists in taking differences between ranging time series in simulations with varying amounts of DM. The resulting residuals (shown in Figure \ref{fig:traj}) are still informative of the results for more sophisticated solar system models because the trajectories of our simulated spacecraft are only majorly affected by the Sun, Jupiter and the planets they come very close to. Additional couplings between other small trajectory perturbations and the gravitational effect of DM are necessarily of higher order. Therefore, their effect vanishes at linear order when considering differences between Doppler time series. When it comes to the prospective mission, this argument is of course only applicable if all the forces acting on the real spacecraft are fully modelled, and the residuals in the time series can be completely attributed to the effect of DM. We are aware that the problem of unmodelled forces, such as the ones that caused the Pioneer anomaly \citep{pioneeranomaly}, will likely be one of the major challenges for achieving the goals we attribute to this mission.
A much larger variation in the final results can be caused by changes in the initial conditions of the spacecraft, specifically in ways that affect their final velocities as they approach the ice giants (as suggested by equation (\ref{eq:badscaling})). In this paper, we selected initial conditions that were roughly compatible with the mission plan detailed in Section \ref{sec:plan}. We are looking forward to revisiting our results with updated initial conditions once an official mission plan is made available to the public.

It is worth mentioning, that many mechanisms exist which could increase the density of DM within the solar system. In particular, the capture of DM by the outer planets could increase densities by a factor anywhere from a few to several hundred times the background halo value \citep{DMcapture2008,DMcapture2009,DMcapture2011}. If the latter case were true, only a marginal improvement from Cassini-era noise would be necessary for a detection, although the reconstruction of the DM profile would be complicated by a more complex radial dependence. Clumpy halo models could also provide a way to increase the density of DM \citep{Moore99,Robert15,clumps2017}, although it would be only a matter of chance if the solar system happened to intersect with a DM substructure during the duration of the mission.

A more complex solar system model could also contextualise the results of our MCMC procedure regarding the standard gravitational parameters. As shown in Figure \ref{fig:corner_mcmc}, we find that a prospective mission could be a precise probe of planetary masses, improving current measurements by several orders of magnitude. If this statement were also true for a more complex solar system model, reducing the Allan deviation of this prospective mission could become relevant for many scientific goals other than probing the dark sector or gravitational waves. It could directly interest the ephemerides and solar system dynamics communities, as well as e.g. the pulsar timing array community, which also can provide precise ephemerides measurements \citep[see e.g.][]{pta1,pta2}. Moreover, it could benefit the solar science community, as the mission could be capable of constraining interesting parameters such as the mass loss rate through winds as a function of radius \citep[see e.g.][]{swind1,swind2}.

\subsection{In situ measurements}
If orbiters are planned, future ice giant missions will likely provide extra-precession measurements that are several orders of magnitude more precise than current astrometric constraints, exceeding the precision of current ephemerides simulations \citep{pitjeva_2013}. If this is the case, any potential discrepancy between a planet's measured and predicted precession rates must be treated carefully, as it is likely being caused by a simple failure of the simulation rather than an exotic effect.
The analysis completed in this paper assumes that, by the 2040s, solar system simulations will also have improved significantly from the current baselines, reducing uncertainties in planetary motion to the degree required for our analysis to be legitimate. While it is unclear how quickly such simulations might evolve in the next twenty years, an intriguing possibility is that the mission itself be used for this purpose, by e.g. measuring planetary masses with unprecedented precision.

More rigorous constraints on MG theories can be obtained by repeating our numerical MCMC simulations with different modifications to the inverse square law rather than a DM density. We adopted the extra precession method in the vein of \cite{sereno_dm}, with the purpose of analysing several different MG scenarios without having to resort to numerical simulations.

\subsection{Prospects for improving the Allan deviation}
The question of how to improve the Allan of deviation ranging missions is discussed extensively in \citet{armstrong}, as well as in \citetalias{soyuer2021} in the context of future ice giant missions. In short, prospects for reducing the main noise sources of radio links are the following:
Antenna mechanical noise can be significantly reduced by using a combination of a smaller and stiffer receiver along with the main dish \citep{armstrong}. Moreover, three point antenna calibration is being currently investigated, and could potentially reduce the Allan deviation of Ka-band ranging missions by one order of magnitude (private conversation with Sami W. Asmar, NASA jet propulsion lab). Tropospheric noise can be reduced by prioritising high altitude facilities, resorting to stratospheric balloons or by increasing the number of simultaneous measurement points \citep{bock98}. An additional option would be to upgrade the Doppler link to a higher frequency band. Recent advances in optical Doppler orbitography show four orders of magnitude improvement in atmospheric phase noise, compared to typical X-band links \citep{optick1,optic0}. While such a link would require more energy to be maintained, it might not be out of the question for future interplanetary missions. The problem of plasma scintillation noise is astrophysical in origin, and therefore harder to deal with. However, it can be reduced by taking the measurements at optimal Sun-Earth-spacecraft configuration, as well as by upgrading the Doppler link to higher frequencies.

We would like to stress that reducing the Allan deviation of Doppler tracking systems is an advantage for many scientific applications other than probing the dark sector. An important objective of in situ observations are precisely measuring the zonal gravity multipole harmonics $J_n$ of the planetary gravity fields, which are an essential for modelling the interior structure and zonal wind dynamics of the ice giants \citep{kaspi2013, benno}. These are also measured by careful reconstruction of the trajectory of orbiters. Furthermore, an improvement in Allan deviation would likely enable the detection of general relativistic effects, such as the frame-dragging due to planetary spins \citep{schaferspin}. An overall improvement of at least two orders of magnitude from Cassini-era values is currently being targeted in the X-band with the purpose of performing precision parametrised post-Newtonian measurements \citep{bepi}. Moreover, it was recently shown in \citetalias{soyuer2021} that improvements to the Allan deviation would lead to the detection of GWs from loud super-massive black hole mergers. Clearly, Doppler tracking of ranging spacecraft is a powerful scientific tool, and further development in noise reduction would benefit several science cases.

\section{Summary \& conclusion}
\label{sec:conclusion}
In Section \ref{sec:ITP}, we have investigated the potential of a prospective ranging mission to the ice giants to improve local constraints on the dark sector. We developed a numerical procedure, based on an MCMC, which simulates the trajectory of ranging spacecraft in various solar system realisations with varying amounts of DM content. The precision of the Doppler link is parametrised by an Allan deviation, scaled on Cassini-era values of $3\times 10^{-15}$. Our results can be summarised as follows:

\begin{itemize}
    \item We estimate that a prospective ice giant mission will be sensitive to DM densities of $ \rho_{\rm{DM}} \sim 9 \times 10^{-20}\, (\sigma_{\rm{A}}/\sigma_{\rm A}^{\rm{Cass}}) $ kg/m$^3$. We show that the $1\sigma$ upper bound $\mathcal{S}_{1\sigma}$ on the expected solar system DM density value of $\rho_{\rm{DM}} = 4.6 \times 10^{-22}\, (\sigma_{\rm{A}}/\sigma_{\rm A}^{\rm{Cass}}) $ kg/m$^3$ is well fitted by the relation $\mathcal{S}_{1\sigma} = 1.0 \times 10^{-20}\, (\sigma_{\rm{A}}/\sigma_{\rm A}^{\rm{Cass}})^{0.8} $ kg/m$^3$.
    
    \item An improvement in Allan deviation by $10^2$ to $10^3$ from Cassini-era values could guarantee the direct detection of DM within our Solar system, with a precision equal to galactic based constraints. This is comparable to the improvement required for the same mission to detect gravitational waves by supermassive black hole binaries and aid space borne gravitational wave detectors in localising sources (see \citetalias{soyuer2021}).
    
\end{itemize}

In Section \ref{sec:ephe}, we have followed the prescription detailed in \citet{sereno_dm, sereno_de}, which consists of transforming measurement uncertainties of the outer planets' extra-precession rates into constraints on the local dark sector. Our results for a prospective ice giant mission can be summarised as follows:
\begin{itemize}  
    \item Ice giant orbiters with good ranging capacity can be used to significantly improve constraints on MG theories. In particular, only a moderate improvement (factor $\sim$10 from Cassini-era) in the Allan deviation is required to rule out Milgrom's original interpolating function \citep{milgrom83}.
    
    \item Any advance in ranging noise reduction guarantees improvements in local constraints on Yukawa-like modifications to the inverse square law. Bounds on the graviton mass could also be improved by several orders of magnitude, out-competing current methods based on the solar system and gravitational waves.
\end{itemize}

Clearly, advances in Doppler tracking can significantly boost the science yield of a prospective ice giant mission, providing a wealth of observations in a wide range of scientific fields. From gravitational waves to dark matter, from the graviton mass to solar winds, we believe that a ranging mission to Uranus and Neptune presents an unique opportunity also for non-planetary science, and that reducing the Allan deviation should become one of the priorities before the scheduled launch in the early 2030s.

\begin{acknowledgements}
    We are grateful to Prasenjit Saha for his firm but gentle leadership, and his spiritual guidance.
      We thank Daniel D'Orazio, Tomas Tamfal, Pedro R. Capelo, Philippe Jetzer and Hugues de Laroussilhe for fruitful discussions. 
\end{acknowledgements}

%
%
\bibliography{ice_giant_dm.bib}
\bibliographystyle{aa}

\appendix

\end{document}